\documentclass[11pt]{article}
\usepackage{hyperref}
\pdfoutput=1
\begin{document}
\title{ It`s Alive! Spontaneous Motion of Shear Thickening Fluids Floating on the Air-Water Interface}
\author{Sunilkumar Khandavalli, Michael Donnell and Jonathan P. Rothstein\\
\\\vspace{6pt} Department of Mechanical and Industrial Engineering
\\ University of Massachusetts, Amherst, MA 01003, USA}
\maketitle

\begin{abstract}
In this fluid dynamics video, we show the spontaneous random motion of thin filaments of a shear-thickening colloidal dispersions floating on the surface of water. The fluid is a dispersion of fumed silica nanoparticles in a low molecular weight polypropylene glycol (PPG) solvent. No external field or force is applied. The observed motion is driven by strong surface tension gradients as the polypropylene glycol slowly diffuses from from the filaments into water, resulting in the observed Marangoni flow. The motion is exaggerated by the thin filament constructs by the attractive interactions between silica nanoparticles and the PPG.
\end{abstract}

\section{Introduction}

Tears of wine, the soap boat, bubble motion are some of the commonly seen amazing effects due to Marangoni flow - stresses generated by surface tension gradient causing fluid motion. Here, we demonstrate a spectacular manifestation of the Marangonic effect by using a shear-thickening fluid. The shear-thickening fluid is hydrophilic fumed silica nanoparticles (AEROSIL@200) in low molecular weight polypropylene glycol (PPG) (M.W 1000 g/mol). These dispersions demonstrate strong shear and extensional thickening behavior. In these two videos, \href{http://ecommons.library.cornell.edu/bitstream/1813/8237/2/LIFTED_H2_EMS
T_FUEL.mpg}{Video1} and \href{http://ecommons.library.cornell.edu/bitstream/1813/8237/4/LIFTED_H2_IEM
_FUEL.mpg}{Video2}, we show a spontaneous and intense erratic motion of filaments of shear-thickening colloidal dispersion at air-water interface. The profound motion which begins as a back and forth motion for long fibres and transitions to high speed spinning as the fibres break and become shorter is due to Marangonic effect - flow driven by stresses resulting from surface tension gradient. The surface tension of PPG is $\sim$ 30 mN/m and that of water is 72.8 mN/m. When PPG is added to the water, the difference in the surface tension drives water away from PPG, due to the resulting local stresses in the fluid. When thin filaments of shear-thickening fluid is added to the water, the Marangoni effect is exaggeraged by the shape of the filament constructs and the slow diffusion of PPG from the filaments which is limited by the presence of nanoparticles and the shear-thickening rheology of the fluid.
%
%
%
\end{document}